\documentclass[runningheads,openright]{svmult}

\usepackage{makeidx}   
\usepackage{graphicx}  
\usepackage{subeqnar}  
\usepackage{multicol}  
\usepackage{physprbb}  
\makeindex             
\input psfig.sty

\newcommand{\age}{\mathrel{\hbox{\rlap{\hbox{\lower4pt\hbox{$\sim$}}}\hbox{$>$}}}}

\newcommand{\eg}{e.g., } 

\newcommand{\etal}{et al.}

\newcommand{\gae}{\mathrel{>\kern-1.0em\lower0.9ex \hbox{$\sim$}}}

\newcommand{\gsim}{\!\!\!\phantom{\ge}\smash{\buildrel{}\over {\lower2.5dd\hbox{$\buildrel{\lower2dd\hbox{$\displaystyle>$}}\over \sim$}}}\,\,}
\newcommand{\gtap}{\mathrel{\hbox{\rlap{\lower.55ex \hbox {$\sim$}}\kern-.3em \raise.4ex \hbox{$>$}}}}
\newcommand{\gtrsim}{\mathrel{\hbox{\rlap{\hbox{\lower4pt\hbox{$\sim$}}}\hbox{$>$}}}}

\newcommand{\ie}{i.e., }

\newcommand{\lae}{\mathrel{<\kern-1.0em\lower0.9ex \hbox{$\sim$}}}
\newcommand{\lesssim}{\mathrel{\hbox{\rlap{\hbox{\lower4pt\hbox{$\sim$}}}\hbox{$<$}}}}
\newcommand{\lsim}{\!\!\!\phantom{\le}\smash{\buildrel{}\over {\lower2.5dd\hbox{$\buildrel{\lower2dd\hbox{$\displaystyle<$}}\over \sim$}}}\,\,}

\newcommand{\ltap}{\mathrel{\hbox{\rlap{\lower.55ex \hbox {$\sim$}} \kern-.3em \raise.4ex \hbox{$<$}}}}
\newcommand{\ltsima}{$\; \buildrel < \over \sim \;$}

\newcommand{\simlt}{\lower.5ex\hbox{\ltsima}} 

\def\CfA{{\sl Harvard-Smithsonian Center for Astrophysics (CfA)}\index{instruments, agencies, observatories, and programs!Harvard-Smithsonian Center for Astrophysics (CfA)}}

\def\CTIO{{\sl Cerro Tololo Inter-American Observatory (CTIO)}\index{instruments, agencies, observatories, and programs!Cerro Tololo Inter-American Observatory (CTIO)}}

\def\E{{\sl Einstein}\index{instruments, agencies, observatories, and programs!Einstein Observatory -- HEAO-2 (Einstein)}}

\def\I{{\sl IRAS}}

\def\LOTOSS{{\sl Lick Observatory and Tenagra Observatory Supernova Search (LOTOSS)}\index{instruments, agencies, observatories, and programs!Lick Observatory!Lick Observatory and Tenagra Observatory Supernova Search (LOTOSS)}}

\def\WFPC2{{\sl Wide-Field Planetary Camera 2 (WFPC2)}\index{instruments, agencies, observatories, and programs!Hubble Space Telescope (HST)!Wide-Field Planetary Camera 2 (WFPC2)}}
\def\WFPc2{{\sl WFPC2}\index{instruments, agencies, observatories, and programs!Hubble Space Telescope (HST)!Wide-Field Planetary Camera 2 (WFPC2)}}

\def\YALO{{\sl Yale-AURA-Lisbon-Ohio (YALO)}\index{instruments, agencies, observatories, and programs!Cerro Tololo Inter-American Observatory (CTIO)!Yale-AURA-Lisbon-Ohio (YALO)}}

\def\CSM{circumstellar medium (CSM)}

\def\NIR{{near infrared (NIR)}\index{GRB!infrared}}
\def\NIr{{NIR}\index{GRB!infrared}}

\def \aap #1 #2 {{Astron. Astrophys.\/} {\bf #1}, #2~}
\def \aar #1 #2 {{Astron. Astrophys. Rev.\/} {\bf #1}, #2~}
\def \aas #1 #2 {{Astron. Astrophys. Suppl. Ser.\/} {\bf #1}, #2~}
\def \aj #1 #2 {{Astron. J.\/} {\bf #1}, #2~}
\def \al #1 #2 {{Astron. Lett.\/} {\bf #1}, #2~}
\def \an #1 #2 {{Astron. Nach.\/} {\bf #1}, #2~}
\def \annap #1 #2 {{Annals Ap.\/} {\bf #1}, #2~}
\def \aph #1 {{astro-ph\/} {#1}~}
\def \ar #1 #2 {{Astron. Rep.\/} {\bf #1}, #2~}
\def \araap #1 #2 {{Ann. Rev. Astron. Astrophys.\/} {\bf #1}, #2~}
\def \asiagoc #1 #2 {{Asiago Contr.\/} {\bf #1}, #2~}
\def \apj #1 #2 {{Astrophys. J.\/} {\bf #1}, #2~}
\def \apjl #1 #2 {{Astrophys. J. Lett.\/} {\bf #1}, #2~}
\def \apjs #1 #2 {{Astrophys. J. Suppl.\/} {\bf #1}, #2~}
\def \apjsub #1 {{Astrophys. J.\/} {#1}~}
\def \apph #1 #2 {{Astropart. Phys.\/} {\bf #1}, #2~}
\def \apss #1 #2 {{Astrophys. Space Sci.\/} {\bf #1}, #2~}
\def \aspc #1 #2 {{ASP Conf.~Proc.\/} {\bf #1}, #2~}
\def \aspl #1 {{ASP Leaflet\/} {#1}~}
\def \asr #1 #2 {{Adv. Space Res.\/} {\bf #1}, #2~}
\def \astrl #1 #2 {{Astron. Lett.\/} {\bf #1}, #2~}
\def \azh #1 #2 {{Astron. Zhurnal\/} {\bf #1}, #2~}
\def \baas #1 #2 {{Bull. Am. Astron. Soc.\/} {\bf #1}, #2~}
\def \ban #1 #2 {{Bull. Astron. Inst. Neth.\/} {\bf #1}, #2~}
\def \basi #1 #2 {{Bull. Astron. Soc. India\/} {\bf #1}, #2~}
\def \ca #1 #2 {{Chinese Astron.\/} {\bf #1}, #2~}
\def \coap #1 #2 {{Contrib. Oss. Astrofis. Padova in Asiago\/} {\bf #1}, #2~}
\def \cap #1 #2 {{Comm. Astrophys.\/} {\bf #1}, #2~}
\def \emsg #1 {{ESO Messenger\/} {#1}~}
\def \gcn #1 {{GCN\/} {#1}~}
\def \hast #1 #2 {{Highlights of Astronomy\/} {\bf #1}, #2~}
\def \iauc #1 {{IAUC\/} {#1}~}
\def \iaus #1 #2 {{IAU Symp. 110: VLBI \& Compact Radio Sources\/} {\bf #1}, #2~}

\def \jcam #1 #2 {{J. Comp. Appl. Math.\/} {\bf #1}, #2~}
\def \jet #1 #2 {{JETP Lett.\/} {\bf #1}, #2~}
\def \jha #1 #2 {{J. Hist. Astron.\/} {\bf #1}, #2~}
\def \jrasc #1 #2 {{J. R. Astron. Soc. Canada\/} {\bf #1}, #2~}
\def \mem #1 #2 {{Mem. R. Astron. Soc.\/} {\bf #1}, #2~}
\def \mess #1 #2 {{The Messenger\/} {\bf #1}, #2~}
\def \mnras #1 #2 {{Mon. Not. R. Astron. Soc.\/} {\bf #1}, #2~}
\def \mplb #1 #2 {{Mod. Phys. Lett. B\/} {\bf #1}, #2~}
\def \nat #1 #2 {{Nature\/} {\bf #1}, #2~}
\def \newa #1 #2 {{New Astron.\/} {\bf #1}, #2~}
\def \nuca #1 #2 {{Nucl. Phys. A\/} {\bf #1}, #2~}
\def \nucb #1 #2 {{Nucl. Phys. B\/} {\bf #1}, #2~}
\def \npps #1 #2 {{Nucl. Phys. Proc. Suppl.\/} {\bf #1}, #2~}
\def \nyasa #1 #2 {{NY Acad. Sci. Ann.\/} {\bf #1}, #2~}
\def \obsy #1 #2 {{The Observatory\/} {\bf #1}, #2~}
\def \phfl #1 #2 {{Phys. Fluids\/} {\bf #1}, #2~}
\def \phytd #1 #2 {{Phys. Today\/} {\bf #1}, #2~}
\def \prl #1 #2 {{Phys. Rev. Lett.\/} {\bf #1}, #2~}
\def \prp #1 #2 {{Phys. Rep.\/} {\bf #1}, #2~}
\def \phyr #1 #2 {{Phys. Rev.\/} {\bf #1}, #2~}
\def \phyrd #1 #2 {{Phys. Rev. D\/} {\bf #1}, #2~}
\def \prasa #1 #2 {{Proc. Astron. Soc. Australia\/} {\bf #1}, #2~}
\def \pasa #1 #2 {{Pub. Astron. Soc. Australia\/} {\bf #1}, #2~}
\def \pasj #1 #2 {{Pub. Astron. Soc. Japan\/} {\bf #1}, #2~}
\def \pasp #1 #2 {{Pub. Astron. Soc. Pacific\/} {\bf #1}, #2~}
\def \qjras #1 #2 {{Q. J. R. Astron. Soc.\/} {\bf #1}, #2~}
\def \rma #1 #2 {{Rev. Mod. Astron.\/} {\bf #1}, #2~}
\def \rpp #1 #2 {{Rep. Prog. Phys.\/} {\bf #1}, #2~}
\def \rpph #1 #2 {{Rev. Plasma Phys.\/} {\bf #1}, #2~}
\def \sait #1 #2 {{Mem.\ Soc.\ Astron.\ It.\/} {\bf #1}, #2~}
\def \sast #1 #2 {{Sov. Astron.\/} {\bf #1}, #2~}
\def \sal #1 #2 {{Sov. Astron. Lett.\/} {\bf #1}, #2~}
\def \sat #1 #2 {{Sky \& Tel.\/} {\bf #1}, #2~}
\def \sci #1 #2 {{Science\/} {\bf #1}, #2~}
\def \spie #1 #2 {{SPIE\/} {\bf #1}, #2~}
\def \shns #1 #2 {{Stud. Hist. Nat. Sci.\/} {\bf #1}, #2~}
\def \va #1 #2 {{Vist. Astron.\/} {\bf #1}, #2~}
\newcommand{\SN}[1]{SN#1\index{supernova!individual!SN#1}}

\newcommand{\SNt}[1]{SN#1\index{supernova!type!SN#1}}
\newcommand{\type}[1]{type #1\index{supernova!type!SN#1}}
\index{agency|see {instruments, agencies, observatories, and programs}}
\index{Baade-Wesselink method|see {supernova, distance, Baade-Wesselink method}}
\index{blastwave|see {(supernova, shock, circumstellar) or (GRB, shock, external)}}
\index{circumstellar shock|see {supernova, shock, circumstellar}}
\index{CLUMPYSYN|see {supernova, model, CLUMPYSYN}}
\index{collapsar|see {GRB, progenitor, collapsar}}
\index{dark burst|see {GRB, optical, dark}}
\index{expanding photosphere method|see {supernova, distance, expanding photosphere method}}
\index{filled-center supernova remnant|see {supernova remnant, plerionic}}
\index{forward shock|see {(supernova, shock, circumstellar) or (GRB, shock, external)}}
\index{GRB!model!interstellar medium interactor|see {GRB, circumburst medium, interstellar medium}}
\index{GRB!model!wind interactor|see {GRB, circumburst medium, stellar wind}}
\index{guest star|see {supernova, historical}}
\index{GRB!progenitor!hypernova|see {hypernova}}
\index{GRB!progenitor!Wolf-Rayet star|see {star, Wolf-Rayet}}
\index{interstellar scintillation|see {GRB, radio, scintillation}}
\index{ML MONTE CARLO|see {supernova, model, ML MONTE CARLO}}
\index{NLTE (non-local thermodynamic equilibrium)|see {supernova, model, NLTE}}
\index{outgoing shock|see {(supernova, shock, circumstellar) or (GRB, shock, external)}}
\index{observatory|see {instruments, agencies, observatories, and programs}}
\index{PHOENIX|see {supernova, model, PHOENIX}}
\index{presupernova wind|see {supernova, progenitor, wind}}
\index{program|see {instruments, agencies, observatories, and programs}}
\index{pulsar|see {star, neutron}}
\index{radio supernova|see {supernova, radio}}
\index{RL NEBULAR|see {supernova, model, RL NEBULAR}}
\index{Sobolev approximation|see {supernova, model, Sobolev approximation}}
\index{supernova!cosmology|see {cosmology}}
\index{synchrotron self-absorption|see {supernova, synchrotron, self-absorption}}
\index{SYNOW|see {supernova, model, SYNOW}}
\index{synthetic spectrum|see {supernova, model, synthetic spectra}}
\index{VLBI|see {instruments, agencies, observatories, and programs, VLBI}}

\begin{document}

\mainmatter

\title*{Optical Light Curves of Supernovae\footnote{to be
published in:
"Supernovae and Gamma Ray Bursters",
Lecture Notes in Physics (http://link.springer.de/series/lnpp)}
}
\toctitle{Optical Light Curves of Supernovae}
%
%
\titlerunning{Optical Light Curves of Supernovae}
%
\author{Bruno Leibundgut\inst{1}
\and Nicholas B.~Suntzeff\inst{2} }
\authorrunning{Leibundgut and Suntzeff}
%
%
\institute{European Southern Observatory, Karl-Schwarzschild-Strasse 2, D-85748 Garching, Germany \and Cerro Tololo Inter-American Observatory, Casilla 603, La Serena, Chile}

\maketitle              

\begin{abstract}
Photometry is the most easily acquired information about
supernovae. The light curves constructed from regular imaging 
provide signatures not only for the energy input, the radiation escape,
the local environment and the progenitor stars, but also for the intervening
dust. They are the main tool for the use of supernovae as distance 
indicators through the determination of the luminosity. 

The light curve of \SN{1987A} still is the richest and longest observed
example for a core-collapse supernova. Despite the peculiar nature of
this object, as explosion of a blue supergiant, it displayed all the
characteristics of \type{II} supernovae. The light curves of \type{Ib/c} 
supernovae are more
homogeneous, but still display the signatures of explosions in
massive stars, among them early interaction with their circumstellar material.

Wrinkles in the near-uniform appearance of thermonuclear (\type{Ia}) supernovae have
emerged during the past decade. Subtle differences have been observed
especially at \NIR\ wavelengths. Interestingly, the light curve
shapes appear to correlate with a variety of other characteristics of
these supernovae.

The construction of bolometric light curves provides the most direct link
to theoretical predictions and can yield sorely needed constraints for
the models. First steps in this direction have been already made. 
\end{abstract}

\section{Physics of Supernova Light Curves}
\label{sec:phys}
The temporal evolution of the energy release by supernovae (SNe) is one of
the major sources of information about the nature of these events. The
brightness information is relatively easy to obtain and, hence, light
curves have been one of the main stays of supernova research. It is not
only the light curves themselves, but also the absolute luminosity and the color
evolution that have provided major insights into the supernova
phenomenon. 
Through light curves it has been possible to distinguish between 
progenitor models, infer some aspects of the progenitor evolution, 
measure the power
sources, detail the explosion models, and probe the local environment 
of the supernova explosions. 

The observational data have been substantially increased over
the last decade (for a status before 1990 see \cite{Kir90}).
In particular, there are now large
databases with light curve data for \type{Ia} supernovae (\SNt{Ia}). The data on \type{II} supernovae has been extensively expanded as well, but there is still
a clear lack of light curves for peculiar objects. Over the last
decade the supernova family has further acquired new members and
subclassifications (see the chapter by Turatto in this volume). 

The energetic display of a supernova can have several different input sources. 
The most important power comes in almost all cases from the radioactive decay of
material newly synthesized in the explosion. The major contributor 
is $^{56}$Ni, the main product of burning to nuclear
statistical equilibrium at the temperatures and densities encountered in
supernovae. This nucleus is unstable and decays with a half-life of
6.1 days due to electron capture to $^{56}$Co emitting $\gamma$-photons
with energies of 750~keV, 812~keV, and 158~keV. 
The cobalt isotope is also unstable and decays with a half-life of 77.1~days
through electron capture
(81\%) and $\beta$-decay (19\%) to $^{56}$Fe. In this process 
$\gamma$-photons with energies of 1.238~MeV and 847~keV are
emitted. The kinetic energy of the electrons is about 600~keV (for more
details on the radioactive decays see \cite{Arn96,Die98,Nad98}). The
$\gamma$-rays are down-scattered or thermalized in the ejecta until they 
emerge as optical or \NIr\ photons \cite{Arn82,Hof95,Pin00-BL}.

The light curves depend on the size and mass of the progenitor star and 
the strength of the explosion.
Additional energy input, which results in modulations of the emerging 
radiation, comes from shock cooling, recombination of the ionized ejecta, 
collision of the shock with \CSM\ and possible accretion onto a compact remnant. 
Light curves are further shaped by the time-variable escape fraction of
$\gamma$-rays, dust formation and
absorption in the interstellar medium. In some cases, forward scattered
light can change the light curves, \eg through light echos and
fluorescence of nearby gas ionized by the X-ray/UV shock breakout of the
explosion.

With this panoply of different energy contributors and modulators, 
light curve displays
are very rich indeed. Despite the plethora of possibilities, light
curves of different supernova types are rather distinct, although not
sufficiently so for a solid classification. They
are, however, important tools to learn about the physics of supernovae.

Light curves are discussed in \cite{Fil97-BL,Kir90,Lei95,Pat93,Pat94}.
Reviews concentrating more specifically on
\SNt{Ia} light curves can be found in \cite{Lei96,Lei00,Mei00,Sun96}, 
while the light curves of core-collapse
supernovae have mostly been summarized in relation to \SN{1987A} \cite{Arn89,McC93}. 
Additional well-sampled data sets are available for
\SN{1993J}, \SN{1998S} and \SN{1999em} (see references below).

The following sections give a brief overview of observational data
sources (\S\ref{sec:obs}), describe the light curves of the main
supernova types (core-collapse supernovae in \S\ref{sec:core} and 
thermonuclear supernovae in
\S\ref{sec:Ia}) and the physics behind the light curve shapes. We will
discuss bolometric light curves in \S\ref{sec:bol} before we summarize
in \S\ref{sec:sum}.

\section{Observations}
\label{sec:obs}

With modern area detectors, light curves of supernovae have become much
easier to assemble. While early light curves have been compiled from
photographic plates \cite{Lei91a,Pat97,Pie95} observations are now recorded 
with CCDs. The increased
sensitivity has allowed astronomers to successfully move to smaller-size
telescopes and to improve the temporal sampling. In parallel, the move
to more robotic telescopes has increased the number of supernova discoveries
tremendously (see, \eg the chapter by Cappellaro in this volume).
There are many observational programs for supernovae
currently in progress (for a description of some see \cite{Lei00}), 
which contribute photometry for many supernovae. Most successful
are efforts with semi-automated telescopes. The robotic telescopes of
the \LOTOSS\ have discovered and followed many supernovae in the last few years 
\cite{Li01a,Ric96}. At the \CTIO\ the \YALO\ telescope has regularly
provided light curves for nearby
supernovae (see, \eg \cite{Kri01,Sun99}). 
light curves have been further contributed by the Padova group (see, \eg \cite{BL-Ben99,Sal01-BL,Tur98a-BL,Tur98b-BL} and the results of the 
\CfA\ team \cite{Jha99-BL,Rie99a}). There are additional contributions on \SN{1993J} 
\cite{Bar95-BL,Ben94,Dor95,Lew94,Mel95,Ric96,Wad97} 
and more recently \SN{1998S} \cite{Fas00} and \SN{1999em} \cite{Ham01-BL,Leo01}.
Many amateur groups are also collecting supernova light curves. These
data are mostly maintained on Web sites (see \cite{Lei00} for a
small collection of sites).

Infrared observations are still rare. A complete compilation of
the available photometry for \SNt{Ia} before 2000 has been provided by Meikle \cite{Mei00}. 
More data are being added (see, \eg \cite{Her00-BL,Kri01}). 
For core-collapse supernovae complete light curves are available only
for \SN{1987A} \cite{Sun90,Sun91},
\SN{1993J} \cite{Wad97}, \SN{1998S} \cite{Fas00}, and
\SN{1999em} \cite{Ham01-BL}.
There are several programs starting up that will concentrate on
\NIr\ light curves with robotic telescopes.

\section{Core Collapse Supernovae}
\label{sec:core}

\begin{figure}[t]
\begin{center}
\includegraphics[width=0.9\textwidth]{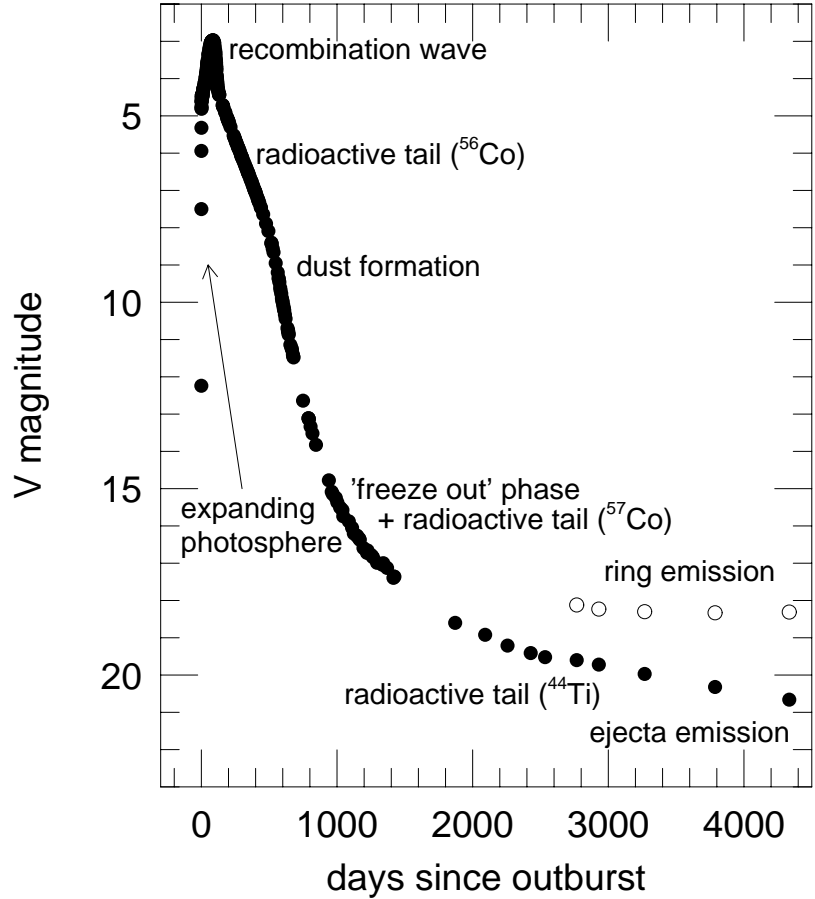}
\end{center}
\caption[]{V light curve of \SN{1987A}. The various phases are
labeled.}
\label{fig:87a}
\end{figure}

Once the shock, which results from the reversal of the core collapse,
breaks out at the surface of the progenitor star the fireworks begin. 
The rapid evolution of 
the core burning just before the collapse is hidden
from the surface due to the long time scales in the atmosphere. The brightness 
of the shock break out is mostly
determined by the temperature in the shock and the size of the progenitor star
\cite{Clo95,Fal77,Kle78}.
This early peak lasts typically from a few hours to a couple of days and
has been observed only for \SN{1987A} \cite{Arn89},
\SN{1993J} \cite{Ric96}, and \SN{1999em} \cite{Ham01-BL,Leo01}.
After a rapid, initial cooling the supernova enters a phase when its
temperature and luminosity remain fairly constant \cite{Eas96,Ham01-BL,Leo01}. 
For supernovae with large 
progenitors the resulting
light curve shows a plateau, while the evolution of supernovae from
smaller stars first exhibits a decline before the supernova brightens
again to reach the plateau.
Examples for the former are \SN{1990E} \cite{Sch93} and \SN{1999em} 
\cite{Ham01-BL,Leo01}, while the \type{II} \SN{1987A} (see, \eg Fig.~\ref{fig:87a}) together 
with the \SNt{Ib/c} belong to
the latter (see, \eg \cite{Lei95}). The plateau originates from a balance
between the receding photosphere in the expanding ejecta \cite{Eas96}. 
During this phase the supernova is powered by the recombination
of hydrogen previously ionized in the supernova shock. The length of the
plateau phase is determined by the depth of the envelope (\ie the envelope
mass and the explosion energy), which is reflected in the expansion velocity 
of the ejecta \cite{Chu91,Pop93}. 
For some objects this plateau phase is conspicuously absent \cite{Pat93}. 
Most prominent among these are \SN{1979C} 
\cite{Bra81} and \SN{1980K} \cite{Bar82}. 

Once the photosphere has receded deep
enough, additional heating from the radioactive decay of $^{56}$Ni and
$^{56}$Co extends the plateau for a brief time \cite{Woo86}. 
Afterwards the light curve is
powered solely by the radioactive decay in the remaining nebula.
The $\gamma$-rays are captured in the ejecta and
converted into optical photons, which can escape freely. At this moment
the supernova light curve drops onto the ``radioactive tail.'' This
happens typically after about 100 days (see Fig.~\ref{fig:87a}).

For a complete trapping of the $\gamma$-rays the luminosity of the late
decline gives an indication of the amount of $^{56}$Ni and $^{56}$Co decays
powering the light curve \cite{Bou91,Sch94a}. 
This can be checked with the decline rate of the bolometric light curve, 
which should reflect the $^{56}$Co decay time. 

From such measurements a rather large range of nickel masses has been
derived \cite{Sch94a,Sol98a-BL,Tur98a-BL}. These phases are especially 
interesting as they may show signatures of 
significant fallback of the inner explosion material onto the forming
compact object, neutron star or black hole, in the explosion 
\cite{Bal00,Ben01-BL,Woo96}.

Especially for \SNt{Ib/c} the decline rates are steeper
\cite{Clo01-BL,Ric96,Sol98b-BL,Sol00} at these times,
which is an indication that some of the $\gamma$-rays escape from the
ejecta without any energy deposition. 

Very few objects have been followed beyond about 200 days and the
situation has not changed very much since about 10 years ago 
\cite{Pat93,Tur90}. The photometry of such objects becomes
very difficult as they fade into the glare of the underlying galaxy. 
The remarkable exception is, of course,
\SN{1987A} on which all very late phase information is based. This supernova
suffered from dust forming within the ejecta, which resulted in an
increase of the decline rate in the optical as light was shifted to the
infrared \cite{Sun92}. This occurred after about 450 days and 
could also be
observed as a shift of emission lines towards the blue as the redder
parts of the lines were absorbed. After about 800 days the light curves
started to flatten again \cite{Sun92} due to energy release of
ionized matter \cite{Fra93}. This so-called
``freeze-out'' stems from tenuous material which was ionized during the
original explosion but recombines on time scales longer than the
expansion time. At later times, the flattening is caused by the energy
input from long-lived $^{57}$Co (half-life of 270 days) and $^{44}$Ti which 
has a half-life of about 60 years \cite{Die98}.

As is apparent from Fig.~\ref{fig:87a}, the very late times are, in fact,
dominated by the emission from the circumstellar inner ring, which was 
ionized by the shock breakout \cite{Fra89-BL}. Around 1500 days
after the explosion the ring emission is stronger than that from the supernova
ejecta itself. 

The closeness of \SN{1987A} has permitted us to
resolve the ring emission and also light echos from interstellar dust
from the supernova ejecta \cite{Sun88,Xu95}. For any 
other supernova these contributions can not be separated and would influence 
the light curve shape. 

Some supernovae do not follow the path described above. They often have a much slower evolution and also display narrow lines in their spectra
(see, \eg \cite{Fil97-BL,Fra02}). These objects remain bright
for a long time and must be powered by a different energy source. The
mostly likely explanation for these objects is interaction of the shock
with a dense \CSM. In this process, kinetic energy is
converted to light and hence an additional energy source can be tapped
for the light curve. Since many of these objects are dominated by line
emission and the line strengths critically depend on density and
composition of both the supernova ejecta and the \CSM,
filter light curves can vary significantly from object to object. A few
classical cases are known so far: \SN{1978K} \cite{Ryd93}, \SN{1986J} 
\cite{Lei91b,Rup87}, \SN{1988Z} \cite{Tur93} and \SN{1995N} \cite{Fra02}.

Of similar nature are probably also objects which can be observed for 
decades after their explosions although at a much lower luminosity. 
The prime examples are \SN{1979C} \cite{FesM93,Van99}, \SN{1980K} 
\cite{Fes90,Fes99-BL,Lei91b}, SN1970G \cite{Fes93}, and \SN{1957D} 
\cite{Lon89,Lon92}. A summary of the
observational characteristics of these objects can be found in Leibundgut 
\cite{Lei94} and the detailed physics of the SN-\CSM\ interaction is given 
in the chapter by Chevalier and Fransson in this volume. It is noteworthy that all these
objects also emit radio waves (see the chapter by Sramek and Weiler in this volume).

On no occasion has it been possible, so far, to observe the emergence of
a pulsar within a supernova. Even for \SN{1987A} the data do not require
any input from a pulsar. Some objects which have been
observed for decades, like \SN{1957D}, \SN{1970G}, \SN{1979C}, \SN{1980K}, still do not require the
energy input corresponding to the expectations from a pulsar powered plerion
(see, \eg \cite{Lei94}). It is more likely that these objects are powered
by interaction of the shock with the \CSM. In all of
these cases the spectrum clearly shows that we are still observing
supernova light and not an underlying stellar association. This is
further supported by changes in the light curves as observed for
\SN{1957D} \cite{Lon92} and \SN{1980K} \cite{Fes99-BL}. 

\section{Thermonuclear Supernovae}
\label{sec:Ia}

The observational situation for \type{Ia} SNe is quite different.
There have been several focused searches for \SNt{Ia}, which have produced
large sets of well-sampled light curves (see, \eg 
\cite{Gol01,Ham96,Her00-BL,Jha99-BL,Li01b,Mei00,Phi99,Rie98a,Rie99a,Sal01-BL,Sun96,Sun99}). 
These data samples have produced a detailed view of \SNt{Ia}.
A recent summary of \SNt{Ia} light curves can be found in Leibundgut 
\cite{Lei00} and Meikle \cite{Mei00}.

The incineration of a
white dwarf, which is the most favored model today (see, \eg \cite{Hil00,Woo86}), 
does not predict an observable shock breaking out at the surface. The rise in brightness
is due to the increase in size of the ejecta and lasts for almost three
weeks \cite{Ald00,Con01,Gol01,Rie99b} with the color and temperature 
rather constant. The earliest observed supernovae are
\SN{1990N} and \SN{1998bu}, which were observed about 17 days before maximum
\cite{Lei00,Rie99b}. The light curves are shaped by the 
progressing diffusion of photons out of the ejecta (see, \eg \cite{Pin00-BL}). 

The maximum is reached first at \NIr\ wavelengths \cite{Con00,Mei00} 
and is followed a few days later in the 
optical. While the blue light curves display a monotonic decrease of the 
brightness for the first month after maximum, the \NIr\ bands 
I, J, H, and K display a prominent second maximum 
after about 20
days for most \SNt{Ia} \cite{Eli81,Mei00}, which is often
also observed as ``shoulders'' in the V and R filter light
curves. This second maximum is conspicuously missing for objects of the 
\type{Ia} faint subclass with \SN{1991bg}
and \SN{1999de} as the most prominent examples \cite{Fil92,Lei93,Mod01-BL,Tur96}.

\SNt{Ia} show a strong color evolution towards the red through the maximum
phase. Despite the difficulty that this poses for an exact
measurement, the intrinsic color of \SNt{Ia} appears to be very uniform
\cite{Phi99} and this is often used to determine the amount of
reddening towards the supernova. 

After about 40 days the light curves settle onto an exponential 
(in luminosity) decline for several months. This has been interpreted as the
optically thin phase when the ejecta nebula captures fewer and fewer
$\gamma$-rays and the optical and \NIr\ light curves decline
faster than the $^{56}$Co decay rate (see, \eg \cite{Hof96-BL,Lei92,Pin00-BL}). 
After about 150 days the
light curves change slope once more (see, \eg \cite{Lei00}) when the 
importance of the positron
channel in the $^{56}$Co decay sets in (see, \eg \S\ref{sec:phys}, 
\cite{Axe80,Mil99-BL}.
The decline at these phases should tell us about the magnetic fields in
the explosion as they determine whether the positrons are captured or
escape the ejecta \cite{Col80,Mil99-BL,Rui98}. Currently, the best 
estimate of these late
light curves predicts a positron escape similar to that of the photons \cite{Mil01-BL}.

Light echos can start to dominate \SNt{Ia} light curves several hundred days
after the explosion. There are now at least two objects with clear signatures of
light echos, \SN{1991T} \cite{Sch94b,Spa99} and
\SN{1998bu} \cite{Cap01}. Their light curves have flattened
almost completely as the peak light is scattered off nearby dust clouds
and, since we observe time and intensity integrated light, the brightness
does not change until the edge of the dust layer is reached
or the scattering angle increases to the point where the scattering
efficiency decreases. High spatial resolution imaging shows rings around
these supernovae (\cite{Spa99} and Garnavich, private communication) very similar to
the ones observed around \SN{1987A}.

Despite the highly complicated hydrodynamics and the radiative
transport in \SNt{Ia} ejecta the bolometric light curves can provide
important insights. Most of the emission from \SNt{Ia} is emitted in the
optical and \NIr\ region \cite{Con01,Sun96}. By
sampling the emission from the atmospheric cutoff near 3600~\AA\ 
to 1~$\mu$m we are capturing about 80\% of the energy emitted by these
objects outside the $\gamma$-ray region. Although the color changes
significantly through the observable life of a \SNt{Ia}, not much light is
emitted in the near-UV (see, \eg \cite{Kir93,Lei96,Sun96}) or the infrared 
\cite{Con01}. In fact, the V
light curve is a rather good surrogate of the bolometric light curve
after maximum \cite{Con00}. In particular, during the
late declines the V light curves have been used to calculate the
Ni mass synthesized in the explosions \cite{Cap98} and also to
estimate the positron escape \cite{Mil01-BL}. 

\begin{figure}[t]
\begin{center}
\includegraphics[width=.9\textwidth]{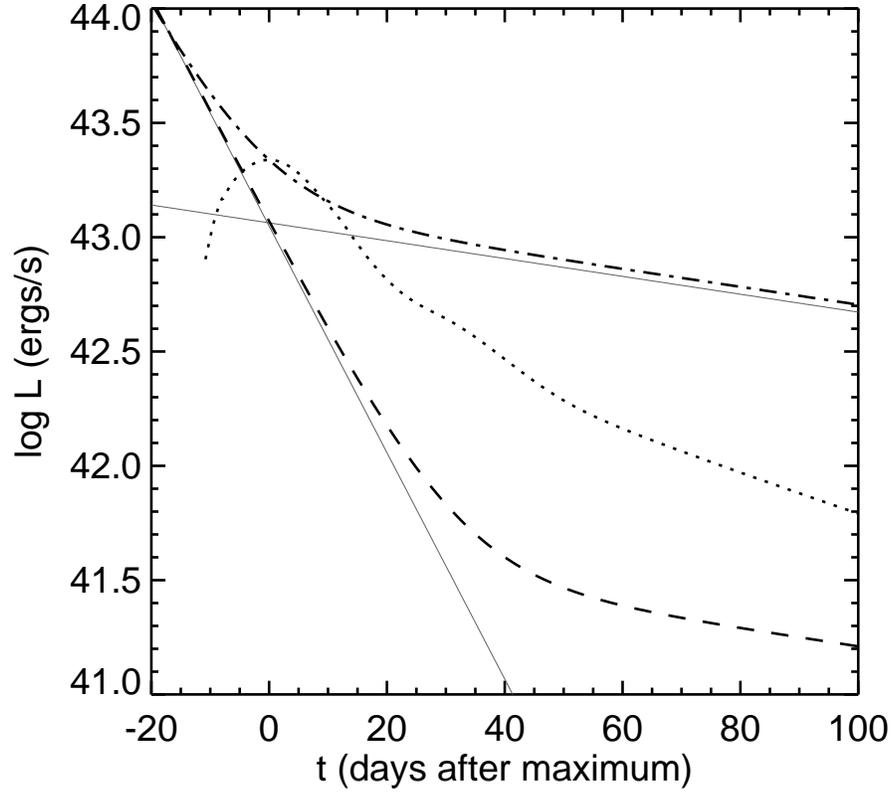}
\end{center}
\caption[]{Bolometric light curve of a \SNt{Ia}. The dotted
line is the observed bolometric light curve. The thin 
lines represent the $^{56}$Ni and $^{56}$Co decay lines, while the
dashed line would be the expected curve if all $\gamma$-rays from the
radioactive decay escape the ejecta and only the positrons are
converted into optical emission.  The dash-dotted line indicates
the expectation of full trapping of all decay energy \cite{Con01}.}
\label{fig:Ia}
\end{figure}

Fig.~\ref{fig:Ia} illustrates the connection between observed bolometric
light curves
(dotted line -- constructed from observations of \SN{1992bc}) and
theoretical models. The expected decay lines for $^{56}$Ni and
$^{56}$Co are indicated by the thin, gray lines. At early times all the
energy is trapped in the ejecta and as the surface increases the
brightness increases as well. Around maximum the energy released is
almost identical to the decay energy from $^{56}$Ni and
$^{56}$Co combined.
This has been pointed out long ago by Arnett \cite{Arn82} and
has been confirmed by Pinto and Eastman \cite{Pin00-BL}. A compilation of typical
values of early models has been given in Branch \cite{Bra92}. 
This is an important feature of
\SNt{Ia} and can be used to measure the total amount of radioactive
material synthesized in the explosion (see, \eg \cite{Con00}). For a
brief time after maximum the energy output from the supernova exceeds the
prediction from a complete trapping of the radioactive energy while
``old'' photons still leak out at the surface. At later phases more and
more $\gamma$-rays are lost and the decline is faster than the line
indicating a full trapping (dash-dotted). The dashed line indicates a
light curve in which no energy from the original $\gamma$ decays is
converted to optical light. Only when 3\% of the $\beta$ decay sets
in can there be some levelling of the light curve. The
observed light curve evolves between these extreme cases.


\subsubsection{Correlations}
\label{sec:Ia_corr}

The \SNt{Ia} light curve shape has been recognized as correlating with
the peak luminosity \cite{Phi93}. This has become the linchpin for
distance determinations using \SNt{Ia} (see, \eg the chapter by Perlmutter and Schmidt in this
volume). The normalization of the peak luminosity allows the
determination of cosmological parameters. However, the correlation is
not as clear-cut as one would wish. There are three implementations
\cite{Per97,Phi99,Rie98a}, which are currently not consistent with each other \cite{Dre00,Lei00}. 

There are other parameters which correlate with the peak luminosity of
\SNt{Ia}. They are the rise time to maximum \cite{Rie99b}, color near 
maximum light \cite{Phi99}, line strengths of the primary Ca and Si 
absorption lines \cite{Nug95-BL}, the velocities as measured in Fe lines 
at late phases \cite{Maz98}, the host galaxy morphology \cite{Sch98}, 
and host galaxy colors \cite{Bra96,Ham95}. There may be indications that the
secondary peak in the I light curves and the shoulder in the
bolometric light curves correlate with the absolute luminosity 
\cite{Con00,Ham96,Rie98a}.

With \SN{2000cx} we now also have a clear example of an object which does
not follow these simple rules. For most objects the rise and decline
rates correlate fairly well \cite{Con00,Rie99b}, but \SN{2000cx} 
violates this rule \cite{Li01b}. While it had a
rapid rise, its light curve did decline more slowly than many
other \SNt{Ia}. It is, at the moment, not clear what fraction of \SNt{Ia} show 
a similar behavior as the rise phase is often not observed and the correlations
have been based on the decline rate. Nevertheless, the fact that for some
\SNt{Ia} the rise and decline rates do correlate \cite{Con00,Rie99b} 
hints at intrinsic differences within this class.

The physical reason for these correlations are not yet clear. Possibilities
are differences in the amount $^{56}$Ni synthesized and the distribution of
the radioactive material, and hence the heating of the ejecta 
\cite{Hof96-BL,Maz01,Pin01}. 

\section{Bolometric Light Curves}
\label{sec:bol}

The connection of the observations to the explosion and radiation physics has to come
through bolometric light curves as they represent the
total energy output of a supernova. This integrated quantity is easily
constructed from the available multi-filter photometry and is also
rather easily extracted from model calculations. Detailed calculations of the
emerging spectrum have proven difficult for all supernova types. While
the physics of the radiation escape for core-collapse supernovae during
the plateau phase, with their large envelopes and a rather well-defined
photosphere (see, \eg the chapter by Branch \etal\ in this volume), is fairly simple to calculate, 
possible chemical abundance mixing and deviations from spherical geometry
can spoil direct comparison with the observations. For
thermonuclear supernovae the spectrum formation is due to multiple
scattering of the $\gamma$-rays in a non-thermal environment and hence
the photons rather ``leak'' out at the surface than originate from a
thermalized photosphere. Hence, the calculation of a filter light curve
requires a detailed spectrum formation calculation which, in the case
of \SNt{Ia}, is currently impossible at the needed detail.

The temporal evolution of the integrated light gives a rather straight
forward comparison between theory and observation. Important explosion
parameters can be calculated from these integrated quantities. Foremost,
the amount of energy available from the nucleosynthesis can be
determined from bolometric light curves. The nickel masses derived for
all supernovae are an essential input of the explosion models. So
far, this observational input has been missing. It is foreseeable that
in the near future this will change and observational constraints on the
models will become available.

The bolometric light curve of \SN{1987A} has been instrumental in decoding
the various phases of the energy release (see  Fig.~\ref{fig:87a} and 
\cite{McC93,Sun92}). The luminosity of the plateau phase indicates
the size of the progenitor and the explosion energy, while the
length of the plateau phase is mostly dominated by the mass of the
exploded star \cite{Pop93}. For the core-collapse supernovae the 
luminosity on the 
radioactive tail, after the light curve leaves the plateau due to the
recombination, together with the time since explosion gives the $^{56}$Ni mass directly. This method has been employed for several supernovae
and it could be shown that there are rather large differences among
these objects. This is particularly interesting as it might provide
insight into fall-back onto the forming neutron star. 

It should be possible to combine some of these measurements into a
picture of the explosion. One can investigate how the progenitor
mass correlates with the explosion energy and the amount of fall-back
onto the forming compact remnant. In extreme cases, it might be possible
to show that a black hole formed in the explosion \cite{Bal00}. 
It is essential to show that the event has all the
signatures of a massive stellar explosion to avoid confusion between
different explosion types. Comparing the explosion parameters with
direct information of the progenitor star, which are becoming available
more often now (see, \eg \cite{Sma01}), will be an important link
between the post-event deductions and the models.

The bolometric light curves of thermonuclear supernovae can be
used to extract physical parameters of the explosions (see, \eg \cite{Vac96}). 
According to Arnett's rule \cite{Arn82,Pin00-BL}, the peak luminosity of
\SNt{Ia} reflects the amount of radioactive $^{56}$Ni produced in the
explosion. This fact has been used by Contardo \etal\ \cite{Con00} (see also
\cite{Con01}) to derive the $^{56}$Ni masses for several supernovae. They
claim that there is a significant range of $^{56}$Ni produced in the
explosions (up to a factor of ten when extreme objects are included). A
similar result was derived by Cappellaro \etal\ \cite{Cap98} based on the late
light curves and the assumption that the V light curve can serve as
a surrogate for the bolometric luminosity. Independent confirmation of
this result comes from the observations of infrared spectra of \SNt{Ia}
\cite{BL-Bow97}. 

The late decline can be further used to determine the escape fraction of
the $\gamma$-rays. A steeper decline indicates a faster decrease of the 
column density. Three possible explanations for this effect could be: 1) different explosion energies, 2) a varying distribution of nickel in the 
explosion, or 3) differences in 
ejecta mass. All possibilities indicate fundamental variations in the
explosions. Interestingly, the expansion velocity of the iron in the
ejecta correlates with peak luminosity of \SNt{Ia} \cite{Maz98},
however not in the way expected from simple models. The brighter
supernovae have larger expansion velocities indicating, for a fixed
ejecta mass, a higher explosion energy. These are also the objects which
produce more $^{56}$Ni in the explosions. In this case, the distribution of
the iron-peak elements must be different to explain
the slower decline at late times \cite{Con01}. An alternative
explanation could be that the ejecta mass for \SNt{Ia} is not the same in all 
events, quite a radical suggestion in view of the currently favored models
of Chandrasekhar-mass white dwarf progenitors (see, \eg \cite{Liv99}).
Bolometric light curves of more objects will be needed to confirm such a
result.


\section{Summary} 
\label{sec:sum}

SN light curves are essential for our understanding of 
supernova physics. The acquisition of light curves has
become relatively easy and the use of robotic telescopes and pipeline
reductions will further advance. The success of recent
supernova searches is increasing the available data set
dramatically and we will soon be able to investigate detailed statistics on light
curve parameters. 

The variety of light curve shapes of core-collapse supernovae provides
evidence for the many physical effects which can influence the outcome
of these events. The rapid, and sometimes violent, evolution of their massive 
progenitor stars and their death within their cradle provides for vastly 
different environments and displays. We can still learn a great deal from
detailed observations. The physics background has developed
considerably over the past decade and the nearby example of \SN{1987A},
offering extremely detailed observations, has led to major new insights
into this phenomenon. 

The short lifetime of core-collapse supernovae can be used to directly
measure the star formation rates as a function of look back time (see, \eg \cite{Dah99} and the chapter by Cappellaro in this volume). However, their light and color
curves are required for a secure identification.

The more uniform appearance of thermonuclear \type{Ia} explosions has been
challenged by the observations in redder filter bands and additional
examples. Although these objects still are rather uniform, their
light curves have allowed us to show that significant variations in these
explosions must exist. The correlation of the light curve shape with the
luminosity has provided a convenient way to normalize these objects and
make them the best cosmological distance indicator available at the
moment (see the chapter by  Perlmutter and Schmidt in this volume). Light curvesare at the
heart of the cosmological applications of \SNt{Ia}. It remains a
major task to establish the physical understanding of these events in the
coming years.



\begin{thebibliography}{100.}
\addcontentsline{toc}{section}{References}

\bibitem{Ald00} G.~Aldering, R.~Knop, P.~Nugent: \aj \textbf{119} 2110 (2000)

\bibitem{Arn82} W.D.~Arnett: \apj \textbf{253} 785 (1982)

\bibitem{Arn96} W.D.~Arnett: \emph{Supernovae and Nucleosynthesis} (Princeton University Press, Princeton 1996)

\bibitem{Arn89} W.D.~Arnett, J.N.~Bahcall, R.P.~Kirshner, S.E.~Woosley: \araap \textbf{27} 629 (1989)

\bibitem{Axe80} T.S.~Axelrod: In: \emph{Type I Supernovae}, ed.~by J.C.~ Wheeler (University of Texas, Austin 1980) p.~80

\bibitem{Bal00} S.~Balberg, L.~Zampieri, S.L.~Shapiro: \apj \textbf{541} 860 (2000)

\bibitem{Bar82} R.~Barbon, F.~Ciatti, L.~Rosino: \aap \textbf{116} 35 (1982)

\bibitem{Bar95-BL} R.~Barbon, S.~Benetti, E.~Cappellaro, F.~Patat, M.~Turatto, T.~Iijima: \aas \textbf{110} 513 (1995)

\bibitem{BL-Ben99} S.~Benetti, M.~Turatto, E.~Cappellaro, I.J.~Danziger, P.A.~Mazzali: \mnras \textbf{305} 811 (1999)

\bibitem{Ben01-BL} S.~Benetti \etal: \mnras \textbf{322} 361 (2001)

\bibitem{Ben94} P.J.~Benson \etal: \aj \textbf{107} 1453 (1994)

\bibitem{Bou91} P.~Bouchet, M.M.~Phillips, N.B.~Suntzeff, C.~Gouiffes.~R.~Hanuschik, D.H.~Wooden: \aap \textbf{245} 490 (1991)

\bibitem{BL-Bow97} E.J.V.~Bowers, W.P.S.~Meikle, T.R.~Geballe, N.A.~Walton, P.A.~Pinto, V.S.~Dhillon, S.B.~Howell, M.K.~Harrop-Allin: \mnras \textbf{290} 663 (1997)

\bibitem{Bra92} D.~Branch: \apj \textbf{392} 35 (1992)

\bibitem{Bra81} D.~Branch, S.W.~Falk, M.L.~McCall, P.~Rybski, A.~Uomoto, B.J.~Wills: \apj \textbf{244} 780 (1981)

\bibitem{Bra96} D.~Branch, W.~Romanishin, E.~Baron: \apjl \textbf{465} L73 (1996)

\bibitem{Cap98} E.~Cappellaro, P.A.~Mazzali, S.~Benetti, I.J.~Danziger, M.~Turatto, M.~Della Valle, F.~Patat: \aap \textbf{328} 203 (1998)

\bibitem{Cap01} E.~Cappellaro \etal: \apjl \textbf{549} L215 (2001)

\bibitem{Chu91} N.N.~Chugai: \sal \textbf{16} 457 (1991)

\bibitem{Clo95} A.~Clocchiatti, J.C.~Wheeler, E.S.~Barker, A.V.~Filippenko, T.~Matheson, J.W.~Liebert: \apj \textbf{446} 167 (1995)

\bibitem{Clo01-BL} A.~Clocchiatti \etal: \apj \textbf{553} 886 (2001)

\bibitem{Col80} S.A.~Colgate, A.G.~Petschek, J.T.~Kriese: \apjl \textbf{237} L81 (1980)

\bibitem{Con01} G.~Contardo: Bolometric light curves of Type Ia Supernovae. PhD Thesis, Technical University, Munich (2001)

\bibitem{Con00} G.~Contardo, B.~Leibundgut, W.D.~Vacca: \aap \textbf{359} 876 (2000)

\bibitem{Dah99} T.~Dahl\'en, C.~Fransson: \aap \textbf{350} 349 (1999)

\bibitem{Die98} R.~Diehl, F.X.~Timmes: \pasp \textbf{110} 637 (1998)

\bibitem{Dor95} V.T.~Doroshenko, Yu.S.~Efimov, N.M.~Shakhovskoi: \astrl \textbf{21} 580 (1995)

\bibitem{Dre00} P.S.~Drell, T.J.~Loredo, I.~Wasserman: \apj \textbf{530} 593 (2000)

\bibitem{Eas96} R.G.~Eastman, B.P.~Schmidt, R.P.~Kirshner: \apj 466, 911 (1996)

\bibitem{Eli81} J.H.~Elias, J.A.~Frogel, J.A.~Hackwell, S.E.~Persson: \apjl \textbf{251} L13 (1981)

\bibitem{Fal77} S.W.~Falk, W.D.~Arnett: \apjs \textbf{33} 515 (1977)

\bibitem{Fas00} A.~Fassia \etal: \mnras \textbf{318} 1093 (2000)

\bibitem{Fes93} R.A.~Fesen: \apjl \textbf{413} L109 (1993)

\bibitem{Fes90} R.A.~Fesen, R.H.~Becker: \apj \textbf{351} 437 (1990)

\bibitem{FesM93} R.A.~Fesen, D.M.~Matonick: \apj \textbf{407} 110 (1993)

\bibitem{Fes99-BL} R.A.~Fesen \etal: \aj \textbf{117} 725 (1999)

\bibitem{Fil97-BL} A.V.~Filippenko: \araap \textbf{35} 309 (1997)

\bibitem{Fil92} A.V.~Filippenko \etal: \aj \textbf{104} 1534 (1992)

\bibitem{Fra89-BL} C.~Fransson, A.~Cassatella, R.~Gilmozzi, R.P.~Kirshner, N.~Panagia, G.~Sonneborn, W.~Wamsteker: \apj \textbf{336} 429 (1989)

\bibitem{Fra93} C.~Fransson, C.~Kozma: \apjl \textbf{408} L25 (1993)

\bibitem{Fra02} C.~Fransson \etal: \apj 572 350 (2002)


\bibitem{Gol01} G.~Goldhaber \etal: \apj \textbf{558} 359 (2001)

\bibitem{Ham95} M.~Hamuy, M.M.~Phillips, J.~Maza, N.B.~Suntzeff, R.A.~Schommer, R.~Avil\'es: \aj \textbf{109} 1 (1995)

\bibitem{Ham96} M.~Hamuy \etal: \aj \textbf{112} 2408 (1996)

\bibitem{Ham01-BL} M.~Hamuy \etal: \apj \textbf{558} 615 (2001)

\bibitem{Her00-BL} M.~Hernandez \etal: \mnras \textbf{319} 223 (2000)

\bibitem{Hil00} W.~Hillebrandt, J.C.~Niemeyer: \araap \textbf{38} 191 (2000)

\bibitem{Hof95} P.~H\"oflich: \apj \textbf{443} 89 (1995)

\bibitem{Hof96-BL} P.~H\"oflich, A.M.~Khokhlov, J.C.~Wheeler, M.M.~Phillips, N.B.~Suntzeff, M.~Hamuy: \apjl \textbf{472} L81 (1996)

\bibitem{Jha99-BL} S.~Jha \etal: \apjs \textbf{125} 73 (1999)

\bibitem{Kir90} R.P.~Kirshner: In: \emph{Supernovae}, ed.\ by A.G.~Petschek (Springer, New York 1990) p.~59

\bibitem{Kir93} R.P.~Kirshner \etal: \apj \textbf{415} 589  (1993)

\bibitem{Kle78} R.I.~Klein, R.A.~Chevalier: \apjl \textbf{223} L109 (1978)

\bibitem{Kri01} K.~Krisciunas \etal: \aj \textbf{122} 1616 (2001)

\bibitem{Lei94} B.~Leibundgut: In: \emph{Circumstellar Media in Late Stages of Stellar Evolution}, ed.~by R.~Clegg, I.~Stevens, P.~Meikle (Cambridge University Press, Cambridge 1994), p.~100

\bibitem{Lei95} B.~Leibundgut: In: \emph{The Lives of Neutron Stars}, ed.~by A.~Alpar, \" U.~Kiziloglu, J.~van Paradijs (Kluwer, Dordrecht 1995) p.~3

\bibitem{Lei96} B.~Leibundgut: In: \emph{IAU Colloquium 145: Supernovae and Supernova Remnants}, ed.~by R.~McCray, (Cambridge University Press, Cambridge 1996) p.~11 

\bibitem{Lei00} B.~Leibundgut: \aar \textbf{10} 179 (2000)

\bibitem{Lei92} B.~Leibundgut, P.A.~Pinto: \apj \textbf{401} 49 (1992)

\bibitem{Lei91a} B.~Leibundgut, G.A.~Tammann, R.~Cadonau, D.~Cerrito: \aas \textbf{89} 537 (1991a)

\bibitem{Lei91b} B.~Leibundgut, R.P.~Kirshner, P.A.~Pinto, M.P.~Rupen, R.C.~Smith, J.E.~Gunn, D.P.~Schneider: \apj \textbf{372} 531 (1991b)

\bibitem{Lei93} B.~Leibundgut \etal: \aj \textbf{105} 301 (1993)

\bibitem{Leo01} D.C.~Leonard \etal: \pasp 114 35 (2002)

\bibitem{Lew94} J.R.~Lewis \etal: \mnras \textbf{266} L27 (1994)

\bibitem{Li01a} W.D.~Li, A.V.~Filippenko, R.R.~Treffers, A.G.~Riess, J.~Hu, Y.~Qiu: \apj \textbf{546} 734 (2001a)

\bibitem{Li01b} W.D.~Li \etal: \pasp \textbf{113} 1178 (2001b)

\bibitem{Liv99} M.~Livio: In: \emph{Type Ia Supernovae: Theory and Cosmology}, ed.~by J.C.~Niemeyer, J.W.~Truran, (Cambridge University Press, Cambridge 1999) p.~33 

\bibitem{Lon89} K.S.~Long, W.P.~Blair, W.~Krzeminski: \apjl \textbf{340} L25 (1989)
\bibitem{Lon92} K.S.~Long, P.F.~Winkler, W.P.~Blair: \apj \textbf{395} 632 (1992)

\bibitem{Maz98} P.A.~Mazzali, E.~Cappellaro, I.J.~Danziger, M.~Turatto, S.~Benetti: \apjl \textbf{499} L49 (1998)

\bibitem{Maz01} P.A.~Mazzali, K.~Nomoto, E.~Cappellaro, T.~Nakamura, H.~Umeda, K.~Iwamoto: \apj \textbf{547} 988 (2001)

\bibitem{McC93} R.~McCray: \araap \textbf{31} 175 (1993)

\bibitem{Mei00} W.P.S.~Meikle: \mnras \textbf{314} 782 (2000)

\bibitem{Mel95} N.V.~Melova, D.Yu.~Tsvetkov, S.Yu.~Shugarov, V.F.~Esipov, N.N.~Pavlyuk: \astrl \textbf{21} 670 (1995)

\bibitem{Mil99-BL} P.A.~Milne, L.-S.~The, M.~Leising: \apjs \textbf{124} 503 (1999)

\bibitem{Mil01-BL} P.A.~Milne, L.-S.~The, M.~Leising: \apj \textbf{559} 1019 (2001)

\bibitem{Mod01-BL} M.~Modjaz \etal \pasp \textbf{113} 308 (2001)

\bibitem{Nad98} D.K.~Nadyozhin: In: \emph{Supernovae and Cosmology}, ed.~by L.~Labhardt, B.~Binggeli, R.~Buser (University of Basel, Basel 1998) p.~125 

\bibitem{Nug95-BL} P.~Nugent, M.M.~Phillips, E.~Baron, D.~Branch, P.~Hauschildt:  \apjl \textbf{455} L147 (1995)

\bibitem{Pat93} F.~Patat, R.~Barbon, R.~Cappellaro, M.~Turatto: \aas \textbf{98} 443 (1993)

\bibitem{Pat94} F.~Patat, R.~Barbon, R.~Cappellaro, M.~Turatto: \aap \textbf{282} 731 (1994)

\bibitem{Pat97} F.~Patat, R.~Barbon, E.~Cappellaro, M.~Turatto: 1997, \aap \textbf{317} 423 (1997)

\bibitem{Per97} S.~Perlmutter \etal: \apj \textbf{483} 565 (1997)

\bibitem{Pie95} M.J.~Pierce, G.H.~Jacoby: \aj \textbf{110} 2885 (1995)

\bibitem{Pin00-BL} P.A.~Pinto, R.G.~Eastman: \apj \textbf{530} 757 (2000)

\bibitem{Pin01} P.A.~Pinto, R.G.~Eastman: \newa \textbf{6} 307 (2001)

\bibitem{Phi93} M.M.~Phillips: \apjl \textbf{413} L105 (1993)

\bibitem{Phi99} M.M.~Phillips, P.~Lira, N.B.~Suntzeff, R.A.~Schommer, M.~Hamuy, J.~Maza: \aj \textbf{118} 1766 (1999)

\bibitem{Pop93} D.V.~Popov: \apj \textbf{414} 712 (1993)


\bibitem{Ric96} M.W.~Richmond, R.R.~Treffers, A.W.~Filippenko, Y.~Paik: \aj \textbf{112} 732 (1996)


\bibitem{Rie98a} A.G.~Riess \etal: \aj \textbf{116} 1009 (1998a)


\bibitem{Rie99a} A.G.~Riess \etal: \aj \textbf{117} 707 (1999a)

\bibitem{Rie99b} A.G.~Riess \etal: \aj \textbf{118} 2675 (1999b)

\bibitem{Rui98} P.~Ruiz-Lapuente, H.C.~Spruit: \apj \textbf{500} 360 (1998)

\bibitem{Rup87} M.P.~Rupen, J.H.~van Gorkom, G.R.~Knapp, J.E, Gunn, D.P.~Schneider: \aj \textbf{94} 61 (1987)

\bibitem{Ryd93} S.~Ryder, L.~Staveley-Smith, M.A.~Dopita, R.~Petre, E.~Colbert, D.~Malin, E.~Schlegel: \apj \textbf{416} 167 (1993)

\bibitem{Sal01-BL} M.E.~Salvo, E.~Cappellaro, P.A.~Mazzali, S.~Benetti, I.J.~Danziger, F.~Patat, M.~Turatto: \mnras \textbf{321} 254 (2001)


\bibitem{Sch93} B.P.~Schmidt \etal: \aj \textbf{105} 2236 (1993)

\bibitem{Sch94a} B.P.~Schmidt \etal: \aj \textbf{107} 1444 (1994a)

\bibitem{Sch94b} B.P.~Schmidt, R.P.~Kirshner, B.~Leibundgut, L.A.~Wells, A.C.~Porter, P.~Ruiz-Lapuente, P.~Challis, A.V.~Filippenko: \apjl \textbf{434} L19 (1994b)

\bibitem{Sch98} B.P.~Schmidt \etal: \apj \textbf{507} 46 (1998)

\bibitem{Sma01} S.J.~Smartt, G.F.~Gilmore, N.~Trentham, C.A.~Tout, C.M.~Frayn: \apjl \textbf{556} L29 (2001)

\bibitem{Sol98a-BL} J.~Sollerman, R.J.~Cumming, P.~Lundqvist: \apj \textbf{493} 933 (1998a)

\bibitem{Sol98b-BL} J.~Sollerman, B.~Leibundgut, J.~Spyromilio: \aap \textbf{337} 207 (1998b)

\bibitem{Sol00} J.~Sollerman, C.~Kozma, C.~Fransson, B.~Leibundgut, P.~Lundqvist, F.~Ryde, P.~Woudt: \apjl \textbf{537} L127 (2000)

\bibitem{Spa99} W.B.~Sparks, F.D.~Maccetto, N.~Panagia, F.R.~Boffi, D.~Branch, M.L.~Hazen, M.~Della Valle: \apj \textbf{523} 585 (1999)

\bibitem{Sun96} N.B.~Suntzeff: In: \emph{IAU Colloquium 145: Supernovae and Supernova Remnants}, ed.~by R.~McCray (Cambridge University Press, Cambridge 1996) p.~41 

\bibitem{Sun90} N.B.~Suntzeff, P.~Bouchet: \aj \textbf{99} 650 (1990)

\bibitem{Sun88} N.B.~Suntzeff, S.~Heathcote, W.G.~Weller, N.~Caldwell, J.P.~Huchra: \nat \textbf{334} 135 (1988)

\bibitem{Sun91} N.B.~Suntzeff, M.M.~Phillips, D.L.~DePoy, J.H.~Elias, A.R.~Walker: \aj \textbf{102} 1118 (1991)

\bibitem{Sun92} N.B.~Suntzeff, M.M.~Phillips, D.L.~DePoy, J.H.~Elias, A.R.~Walker: \apjl \textbf{384} L33 (1992)

\bibitem{Sun99} N.B.~Suntzeff \etal: \aj \textbf{117} 1175 (1999)

\bibitem{Tur90} M.~Turatto, E.~Cappellaro, R.~Barbon, M.~Della Valle, S.~Ortolani, L.~Rosino: \aj \textbf{100} 771 (1990)

\bibitem{Tur93} M.~Turatto, E.~Cappellaro, I.J.~Danziger, S.~Benetti, C.~Gouiffes, M.~Della Valle: \mnras \textbf{262} 128 (1993)

\bibitem{Tur96} M.~Turatto, S.~Benetti, E.~Cappellaro, I.J.~Danziger, M.~Della Valle, C.~Gouiffes, P.A.~Mazzali, F.~Patat: \mnras \textbf{283} 1 (1996)

\bibitem{Tur98a-BL} M.~Turatto \etal: \apjl \textbf{498} L129 (1998)

\bibitem{Tur98b-BL} M.~Turatto, A.~Piemonte, S.~Benetti, E.~Cappellaro, P.A.~Mazzali, I.J.~Danziger, F.~Patat: \aj \textbf{116} 2431 (1998)

\bibitem{Vac96} W.D.~Vacca, B.~Leibundgut: \apjl \textbf{471} L37 (1996)

\bibitem{Van99} S.D.~Van Dyk \etal: \pasp \textbf{111} 313 (1999)

\bibitem{Wad97} T.~Wada, M.~Ueno: \aj \textbf{113} 231 (1997)

\bibitem{Whe00-BL} J.C.~Wheeler, S.~Benetti: In: \emph{Allen's Astrophysical Quantities, 4th edition}, ed.~by A.N.~Cox (AIP Press, New York 2000) p.~451

\bibitem{Woo96} S.W.~Woosley, F.X.~Timmes: 1996, \nuca \textbf{606} 137 (1996)

\bibitem{Woo86} S.E.~Woosley, T.A.~Weaver: \araap \textbf{24} 205 (1986)

\bibitem{Xu95} J.~Xu, A.P.S.~Crotts, W.E.~Kunkel: \apj \textbf{451} 806 (1995)

\end{thebibliography}
\end{document}